\documentclass[%
 reprint,
superscriptaddress,
 amsmath,amssymb,
 prl,
]{revtex4-1}

\usepackage{graphicx}% Include figure files
\usepackage{dcolumn}% Align table columns on decimal point
\usepackage{bm}% bold math
\begin{document}

\preprint{APS/123-QED}

\title{Kinetic Interpretation of Resonance Phenomena in Low\\ Pressure Capacitively Coupled Radio Frequency Plasmas}% Force line breaks with \\

\author{S. Wilczek}
\author{J. Trieschmann}
\author{D. Eremin}
\author{R. P. Brinkmann}

\affiliation{%
Department of Electrical Engineering and Information Science, Ruhr University Bochum, Bochum, Germany
 }%


\author{J. Schulze}
\author{E. Schuengel}
\affiliation{Department of Physics, West Virginia University, Morgantown, WV 26506, USA}
 %This line break forced}%
%\affiliation{Third institution, the second for Charlie Author}

\author{A. Derzsi}
\author{I. Korolov}
\author{P. Hartmann}
\author{Z. Donk\'o}
\affiliation{%
Institute for Solid State Physics and Optics, Wigner Research Centre for Physics, Hungarian Academy of Sciences, 1121 Budapest, Konkoly Thege Mikl\'os str. 29-33, Hungary
}%
\author{T. Mussenbrock}
\affiliation{%
	Department of Electrical Engineering and Information Science, Ruhr University Bochum, Bochum, Germany
}%


\date{\today}% It is always \today, today,

\begin{abstract}
\noindent
The kinetic origin of resonance phenomena in capacitively coupled radio frequency plasmas is discovered based on particle-based numerical simulations. The analysis of the spatio-temporal distributions of plasma parameters such as the densities of hot and cold electrons, as well as the conduction and displacement currents reveals the mechanism of the formation of multiple electron beams during sheath expansion. The interplay between highly energetic beam electrons and low energetic bulk electrons is identified as the physical origin of the excitation of harmonics in the current.
%\begin{description}
%\end{description}
\end{abstract}

\pacs{Valid PACS appear here}% PACS, the Physics and Astronomy
\maketitle

%\tableofcontents
\noindent
%Low temperature plasmas are irreplaceable tools for etching and deposition processes in microelectronics production \cite{LiebermanBook,ChabertBook}. At the same time, they are challenging physical systems, due to their non-linear nature and manifold physical processes. There is a strong demand for the optimization of process control and performance based on a detailed understanding of the plasma physics. A central issue is the understanding of the complex particle heating dynamics on a nanosecond timescale within a radio frequency (RF) period and the development of methods for its control.\\
%\cite{Roadmap1}.\\

\noindent
Capacitively coupled radio frequency (CCRF) discharges are indispensable tools for
semiconductor manufacturing and other innovative applications\cite{LiebermanBook,ChabertBook}. At the same time,
they are challenging physical systems due to their complex and nonlinear dynamics.
At low neutral gas pressures of a few Pa or less, CCRF discharges are operated in a strongly non-local regime. In the so-called ``$\alpha$-mode'', electron heating is dominated by stochastic sheath expansion heating \cite{Belenguer1990} and electric field reversal during sheath collapse \cite{VenderFieldRev,TurnerFieldRev,FieldReversalSchulze,UCZFieldRev}. Stochastic heating was modelled extensively in the past in the frame of a hard wall model, as well as pressure heating \cite{HWModel2,HWModel,PrHeat1,PrHeat2,PrHeat3,Turner1,Kaganovich1,Kaganovich2,Lafleur1,Lafleur2}. During the phase of sheath expansion, energetic electron beams are generated and propagate into the plasma bulk, where they sustain the discharge via ionization and lead to a Bi-Maxwellian electron energy distribution function (EEDF) \cite{WoodBeams,FTCBeams,Schulze1,Schulze2,AmbHeat,BRH1,BRH2}. \\
\noindent
At low pressures, resonance effects such as the plasma series resonance (PSR) \cite{MussenbrockPSR,LiebermanPSR,Mussenbrock,Mussenbrock2} and the plasma parallel resonance (PPR) \cite{Annaratone,Ku,Ku2} can be self-excited and strongly enhance the electron heating \cite{FTCBeams,Mussenbrock}. In the presence of a sinusoidal driving voltage waveform, the excitation of the PSR results in a non-sinusoidal RF current, due to the appearance of harmonics of the driving frequency \cite{Mussenbrock2}. Although these have been observed in experiments \cite{Schulze2}, they are usually neglected in most models of electron heating in CCRF plasmas. Existing theories which include resonance effects are zero-dimensional global models based on equivalent electrical circuits \cite{Mussenbrock} as well as spatially resolved models based on the cold plasma approximation \cite{Mussenbrock2}. As these models do not include any kinetic effects, such resonances should be investigated on a microscopic kinetic level.
A kinetic interpretation is required to clarify some of the most important open questions about electron heating dynamics in CCRF plasmas: What is the kinetic origin of the generation of high frequency (HF) oscillations of the RF current and the generation of multiple electron beams during one phase of sheath expansion such as observed in previous works \cite{Wilczek}? In what way is current continuity ($\nabla \cdot \vec{j}_{\rm tot} = 0$) ensured at all times within the RF period in the presence of electron beams, where the total current density $\vec{j}_{\rm tot} =  \vec{j}_{d} +  \vec{j}_{c}$ is decomposed into the displacement and conduction current density?\\
Our aim is to provide access to a kinetic interpretation of resonance phenomena in CCRF discharges based on Particle-in-Cell simulations complemented by Monte-Carlo treatment of collision processes (PIC/MCC). We consider a single frequency ($f$ = 55 MHz) low pressure ($p$ = 1.3 Pa) argon plasma driven capacitively by a sinusoidal voltage source. The simulation code is 1-dimensional in space and 3-dimensional in velocity space. The electrodes are assumed to be infinite, planar, parallel, and separated by a gap of $L$ = 1.5 cm. One of the electrodes (at $x$ = 0) is driven by a sinusoidal voltage waveform, $\phi(t) = \phi_0 \sin{(2 \omega_{\rm RF} t)}$, where $\omega_{\rm RF} = 2 \pi \cdot 55$ MHz and the other electrode ($x=L$) is grounded. A symmetrical setup is adopted to demonstrate that resonance phenomena are of general relevance even in symmetric discharges, where they are usually neglected. First, we analyze the discharge behavior at $\phi_0 = 150$ V driving voltage amplitude and subsequently, we discuss the discharge behavior at $\phi_0 = 300$ V. Secondary electron emission and particle reflection at the electrodes are neglected in order to simplify the interpretation of the electron heating dynamics. The cross sections for electron-atom and ion-atom collisions are taken from \cite{Phelps,sim3}. For more details see \cite{Wilczek}. 

\begin{figure}
	\begin{center}
		\includegraphics[width=0.5\textwidth]{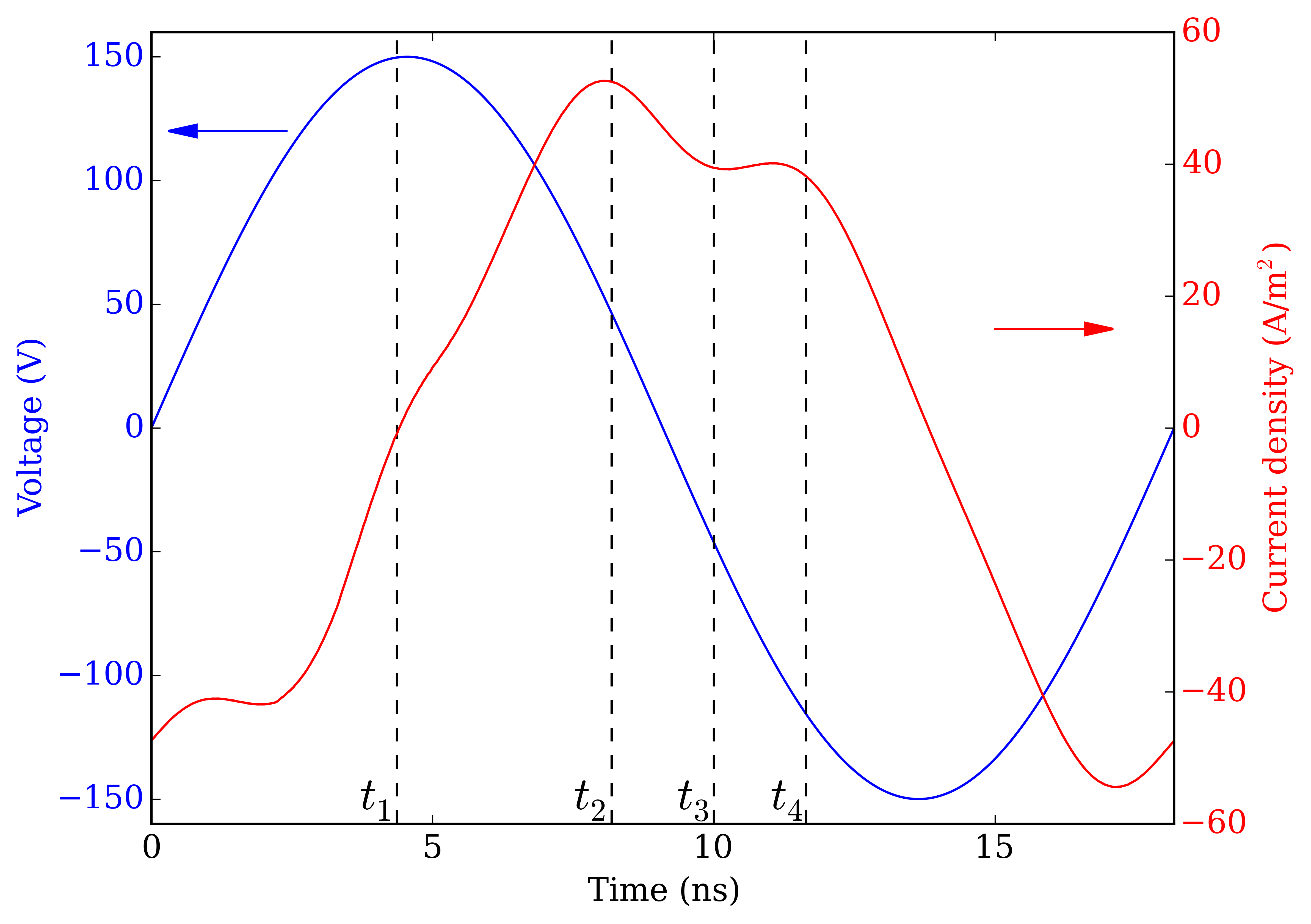}
		\caption{Driving voltage waveform $\phi_0 = 150$ V (left scale) and calculated current waveform (right scale) for one RF period. The vertical dashed lines indicate reference times ($t_1,  t_2, t_3, t_4$) used in the forthcoming analysis.}
		\label{VI}
	\end{center}
\end{figure}

\noindent
Figure \ref{VI} shows the applied sinusoidal driving voltage and the calculated current waveform that is found to be nearly 90 degrees phase shifted and non-sinusoidal as a consequence of the self-excitation of the PSR \cite{Mussenbrock,Mussenbrock2,FTCBeams}. The vertical dashed lines indicate characteristic reference times which are used in the following discussion. 
We note that due to their inertia ($\omega_{\rm pi} \ll \omega_{\rm RF}$, where $\omega_{\rm pi}$ is the ion plasma frequency) the motion of ions is not modulated with the excitation frequency. Ions follow the time-averaged electric field.

\begin{figure}[h!]
	\begin{center}
		
			\includegraphics[width=0.5\textwidth]{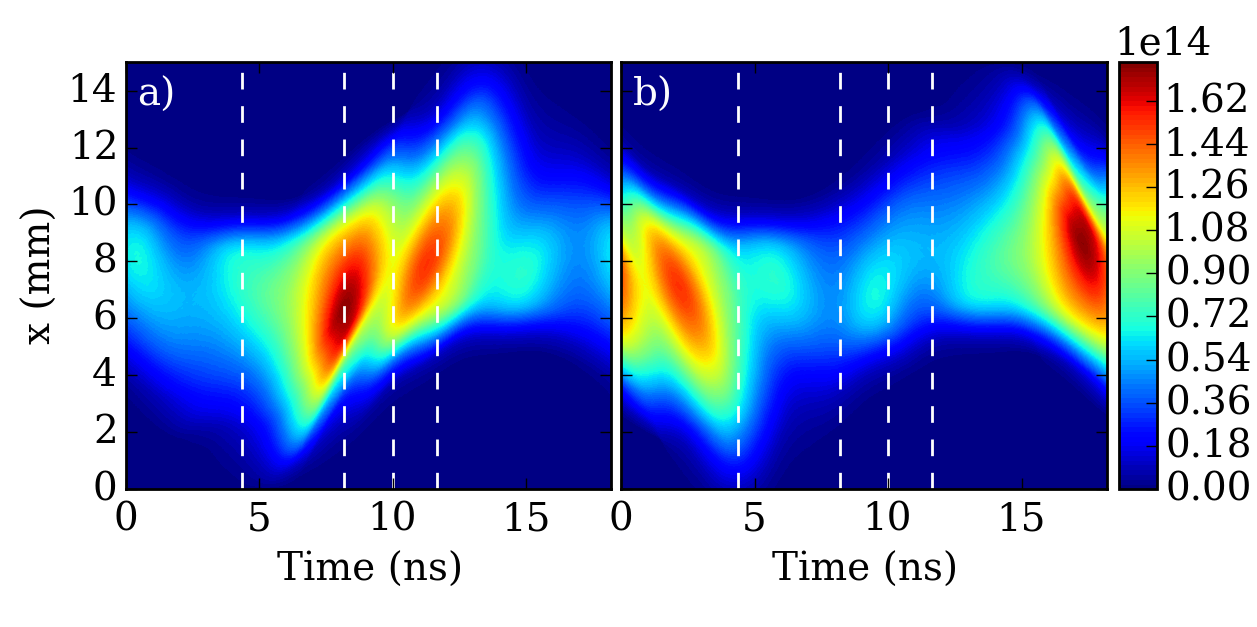}\\
			\includegraphics[width=0.5\textwidth]{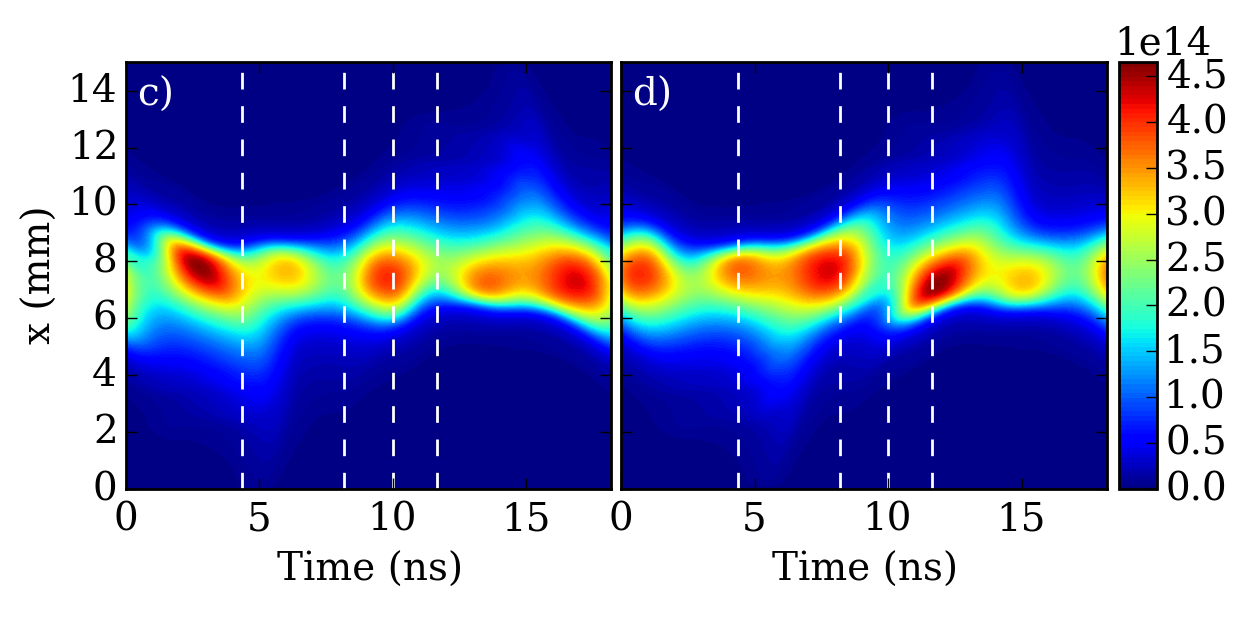}
			
		\caption{Top panels: Density of hot electrons ($\varepsilon >$ 11 eV) that move upwards (a) and downwards (b). Bottom panels: Density of cold electrons below 4 eV that move downwards (c) and upwards (d). The time axis covers one RF period. Densities are provided in m$^{-3}$. The vertical dashed lines indicate the reference times specified in Figure 1. The powered electrode (150 V) is at the bottom (at $x=0$), while the grounded electrode is at the top (at $x=L$).}
		\label{Ext}
	\end{center}
\end{figure}

\begin{figure}[h!]
	\begin{center}
		
		\includegraphics[width=0.5\textwidth]{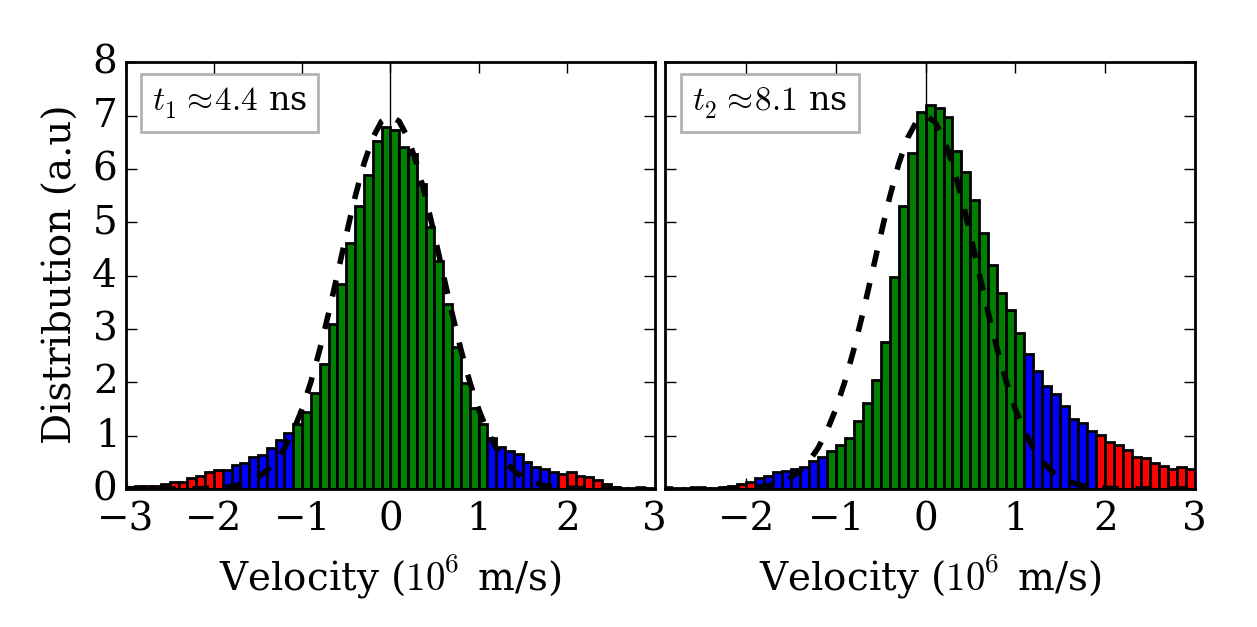}\\
		\includegraphics[width=0.5\textwidth]{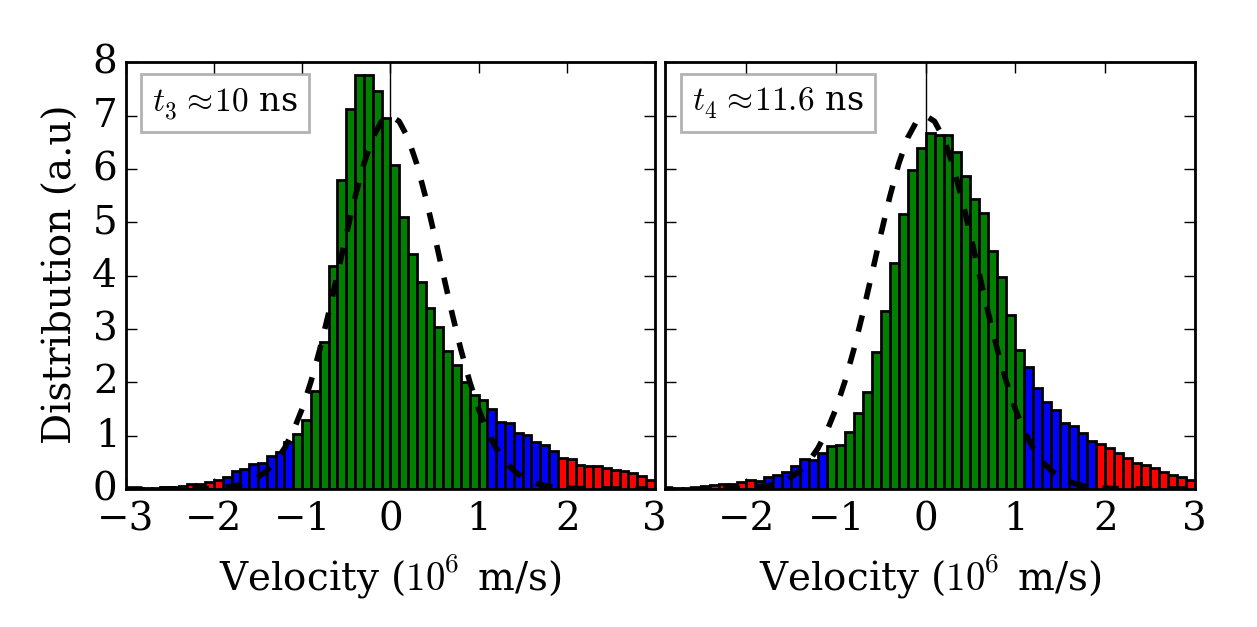}
		
		\caption{The momentary electron velocity distribution function (EVDF) at the considered reference times, spatially averaged over the plasma bulk (6 mm $\le$ x $\le$ 9 mm). The dashed lines indicate a Maxwellian distribution, which is fitted for $t = t_1$. Electrons above 11 eV are represented by red bars, electrons below 4 eV by green bars and electrons between 4 and 11 eV by blue bars ($\phi_0 = 150$ V).}
		\label{Dist}
	\end{center}
\end{figure}

In order to unveil the kinetic origin of the non-linear current waveform, it is necessary to separate the dynamics of cold and hot electrons. The energy threshold for hot electrons is taken to be $\varepsilon_{\rm h}$ = 11 eV, which is in the vicinity of the first excitation level of Ar atoms, while cold electrons are defined to have an energy below $\varepsilon_{\rm c}$ = 4 eV. The energy is calculated using all three velocity components ($\varepsilon_{h,c} = 0.5 m_{e}(v^2_{x} + v^2_{y} + v^2_{z})$). Figure \ref{Ext} shows the density of electrons partitioned according to their energy (as explained above) and according to their direction of velocity. Electrons with $v_x > 0$ (i.e., those moving from the powered electrode towards the grounded electrode) are defined to move ``upwards'', while electrons with $v_x < 0$ are defined to move ``downwards'', in accordance with the representation of the spatio-temporal distributions displayed in Figure \ref{Ext}. The first row of Figure \ref{Ext} shows spatio-temporal plots of the number densities of hot electrons with energies above 11 eV only, which move upwards [Fig. \ref{Ext} (a)] and downwards [Fig. \ref{Ext} (b)]. Similarly to previous works \cite{Wilczek,FTCBeams}, the generation of multiple beams of energetic electrons is observed adjacent to each electrode during the phase of sheath expansion [here: 2 pronounced and 1 weak beam, Figs. \ref{Ext} (a) and (b)]. The bottom row presents the density of cold electrons with energies below 4 eV.
These results are in strong contrast to the prevailing picture of the generation of a single electron beam during one sheath expansion phase. The generation of multiple electron beams strongly affects the spatio-temporal ionization and excitation dynamics, which show similar structures \cite{Wilczek} (i.e., this phenomenon is essential for the generation of such plasmas). Figure \ref{Ext} includes the same vertical lines as Figure \ref{VI} indicating the same characteristic times within one RF period. 
Figure \ref{Dist} shows the momentary electron velocity distribution function (EVDF) at these times, spatially averaged over the plasma bulk [6 mm $\le$ x $\le$ 9 mm, Figs. \ref{Dist} (a) - (d)]. In each plot, electrons with energies above 11 eV are represented by red bars, electrons with energies below 4 eV are marked in green and electrons with energies between 4 eV and 11 eV are marked in blue. Here, the energy is calculated using only the $v_x$-component (axial direction), since the $v_y$ and $v_z$ components do not couple due to the fact that electrons are accelerated only in axial direction by the expanding sheaths and almost no collisions cause an angular scattering of electron beams. The dashed lines in Fig. \ref{Dist} correspond to a Maxwellian distribution fitted to the data shown in Fig. \ref{Dist} (a). 

In the following, a detailed kinetic interpretation of the electron heating dynamics during the phase of sheath expansion at the bottom electrode is presented. (Due to the symmetry, the phenomena occur at the top electrode half an RF period later.) During sheath collapse ($t_1 \approx 4.4$ ns) at the bottom electrode the total current is zero and the EVDF is approximately Maxwellian. Following the start of the sheath expansion ($t_2 \approx 8.1$ ns), the first beam of energetic electrons is generated and propagates towards the plasma bulk [Fig. \ref{Ext}]. These beam electrons move away from the expanding sheath edge leading to a strongly anisotropic EVDF [Fig. \ref{Dist} ($t_2$)]. Clearly, the number of electrons at high positive velocities is increased. As the beam electrons move away from the expanding sheath edge, they leave a positive space charge behind. This positive space charge causes an electric field that accelerates electrons back towards the sheath edge. Consequently, shortly after ($t_3 \approx 10$ ns) cold bulk electrons move back towards the sheath edge [Fig \ref{Ext} ($t_3$)]. At this time, energetic beam electrons move upwards, while cold bulk electrons move downwards [see Fig. \ref{Dist} (c)] (i.e., two groups of electrons move into opposite directions simultaneously). The bulk electrons cannot respond instantaneously to the perturbation caused by the energetic beam electrons due to their inertia. They can only respond on the timescale of the local electron plasma frequency ($\omega_{\rm pe} \approx 5 \omega_{\rm RF}$ in the bulk), which is approximately $3.5$ ns ($\tau \approx 2\pi/ \omega_{\rm pe}$). This leads to a modulation of the electron density in the bulk at the local electron plasma frequency. (Recall that the inertia of electrons is the kinetic reason why an ``inductance'' is required in global equivalent circuit models to excite resonance effects.) Later ($t_4 \approx 11.6$ ns), the drifting cold bulk electrons approach the energy barrier of the expanding sheath edge and upon impact a second beam of energetic electrons is generated [Fig. \ref{Dist} ($t_4$)]. This process is repeated until the sheath expansion stops. Under these conditions, this mechanism leads to the generation of two pronounced energetic electron beams during one phase of sheath expansion at $t \approx$ 8.1 ns, 11.6 ns. Furthermore, a third weak beam formation at 15 ns is present. For the latter the modulated bulk electrons reach the end of the sheath expansion and, therefore, experience just a slight kick by the sheath potential.
This kinetic picture links the complex dynamics of the electrons to the non-linearity of the RF current.  

\begin{figure}[h!]
	\begin{center}
			\includegraphics[width=0.50\textwidth]{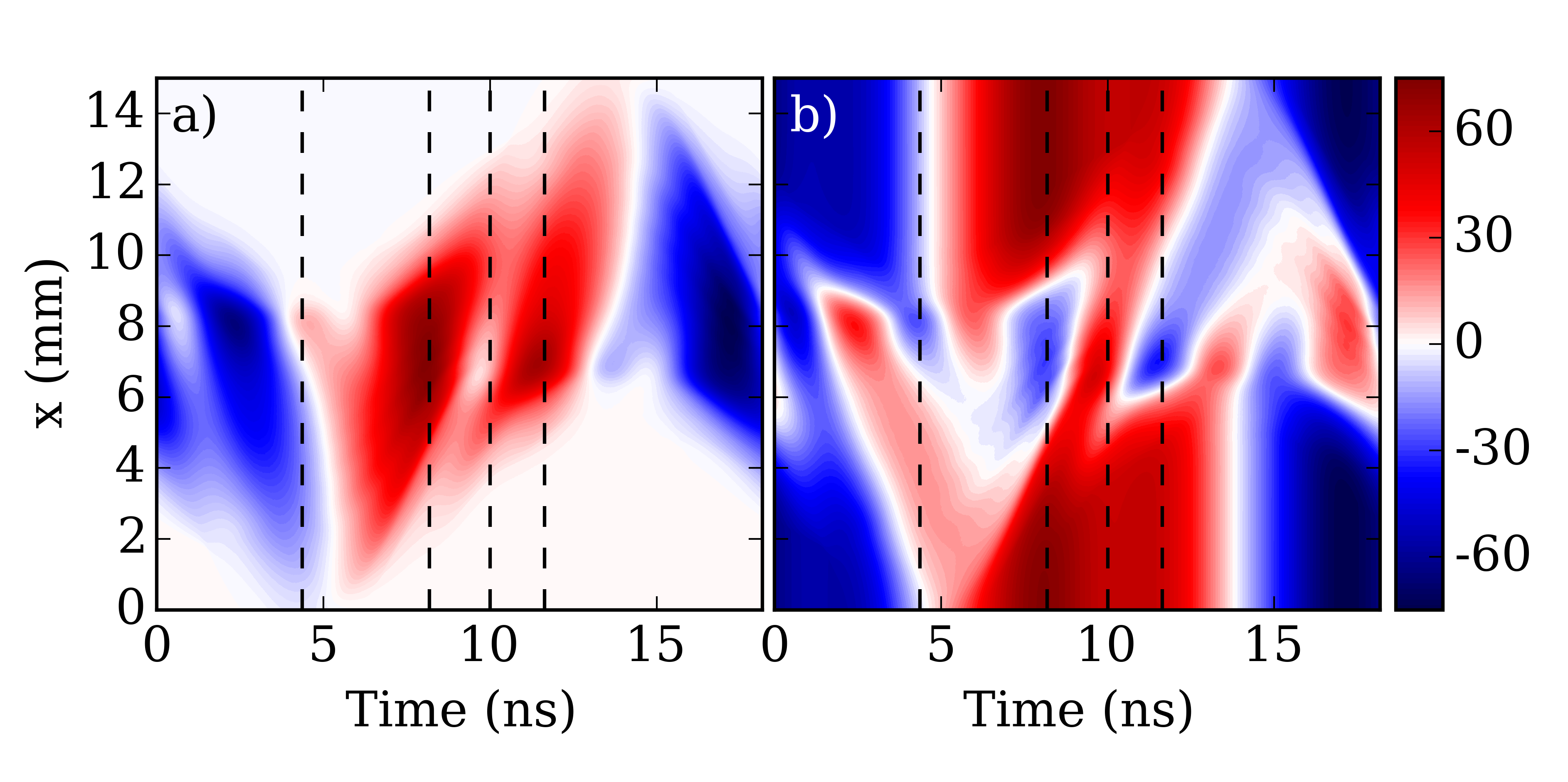}
			
		\caption{Spatio-temporal distribution of the conduction (a) and displacement (b) current density for one RF-cycle in A/m$^2$ ($\phi_0 = 150$ V).}
		\label{Curr}
	\end{center}
\end{figure}

\begin{figure}[h!]
	\begin{center}
		\includegraphics[width=0.5\textwidth]{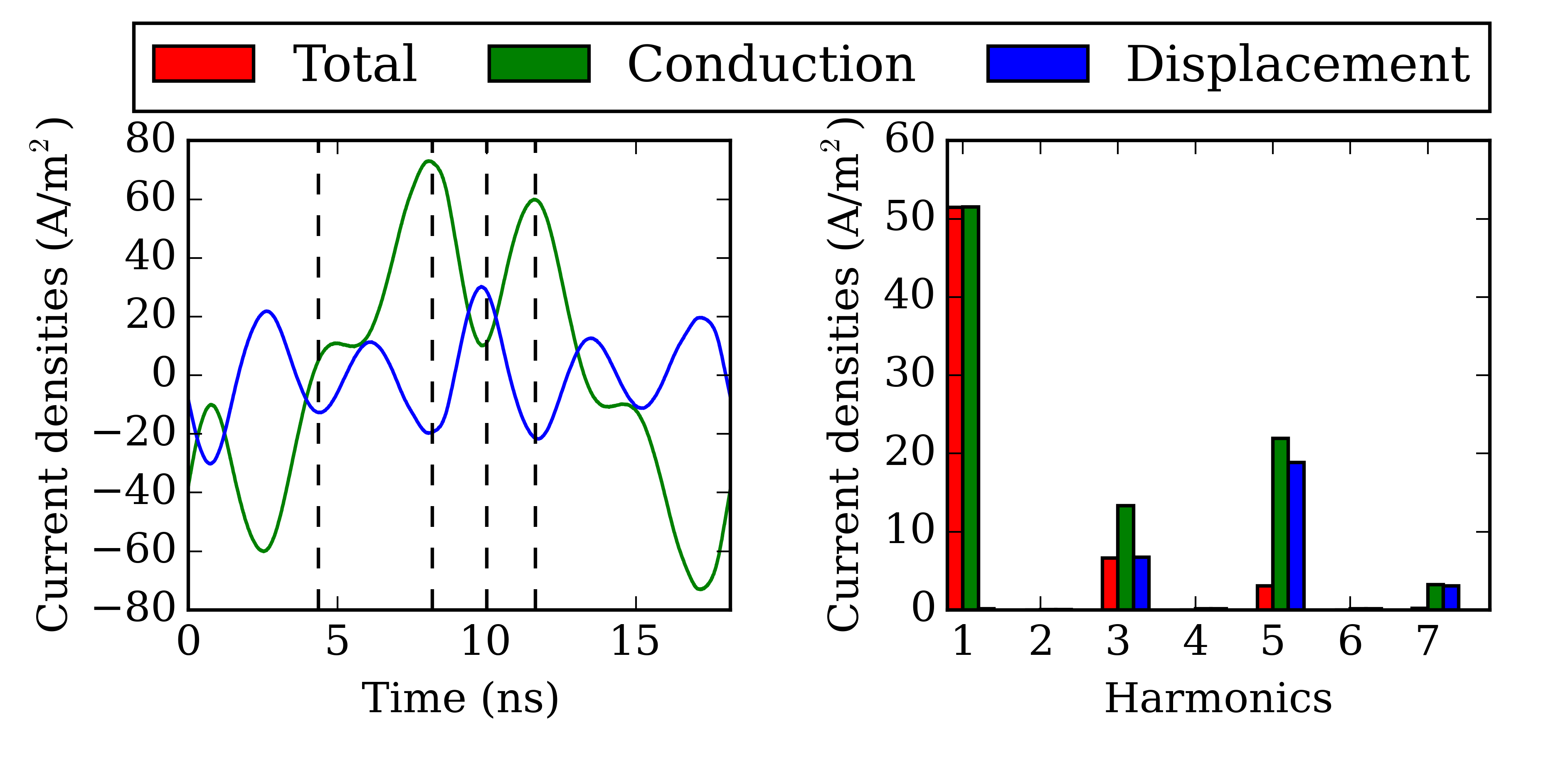}
		\\
		\caption{The conduction and displacement current density (left) for one RF-cycle and the corresponding Fourier spectra (right) in the center, at 7.5 mm ($\phi_0 = 150$ V).}
		\label{Curr2}
	\end{center}
\end{figure}

\begin{figure}[h!]
	\begin{center}
			\includegraphics[width=0.5\textwidth]{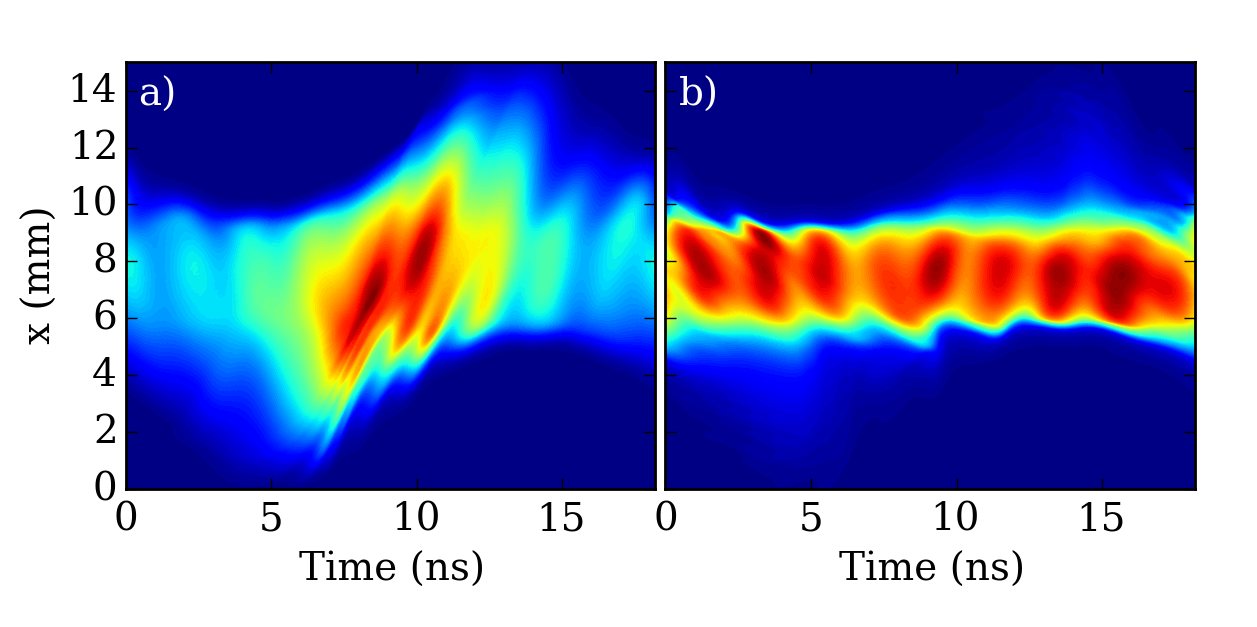}
			
		\caption{Density of hot electrons ($\varepsilon >$ 11 eV) that move upwards (a) and density of cold electrons below 4 eV that move downwards (b) in m$^{-3}$ ($\phi_0 = 300$ V). Similar color scale as shown in Fig. \ref{Ext} from 0 to a) 4.05 10$^{14}$ and b) 1.21 10$^{15}$ m$^{-3}$.}
		\label{ele300}
	\end{center}
\end{figure}

\begin{figure}
	\begin{center}
		\includegraphics[width=0.5\textwidth]{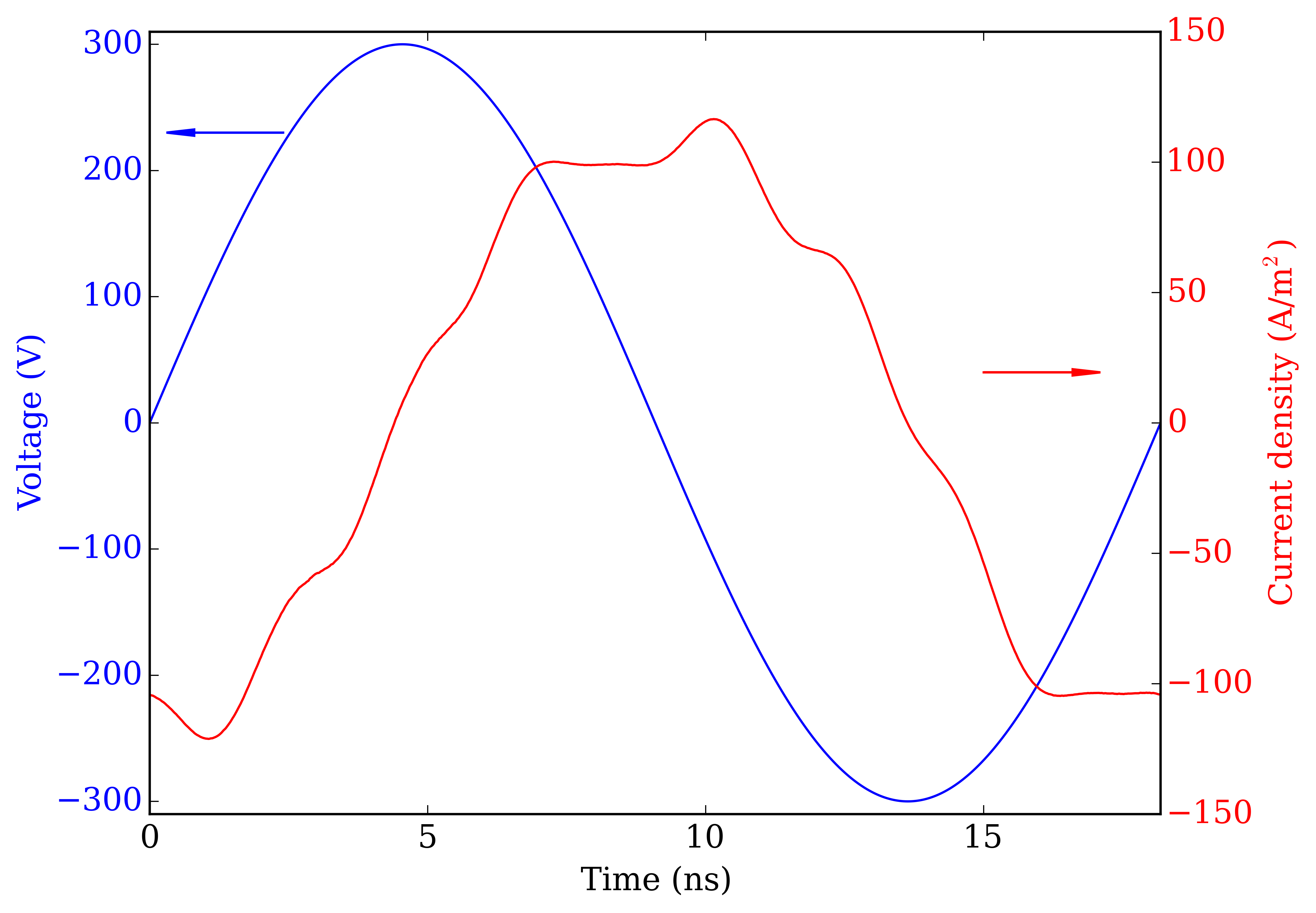}
		\caption{Driving voltage waveform $\phi_0 = 300$ V (left scale) and calculated current waveform (right scale) for one RF period.}
		\label{VI_300V}
	\end{center}
\end{figure}

\begin{figure}[h!]
	\begin{center}

			\includegraphics[width=0.5\textwidth]{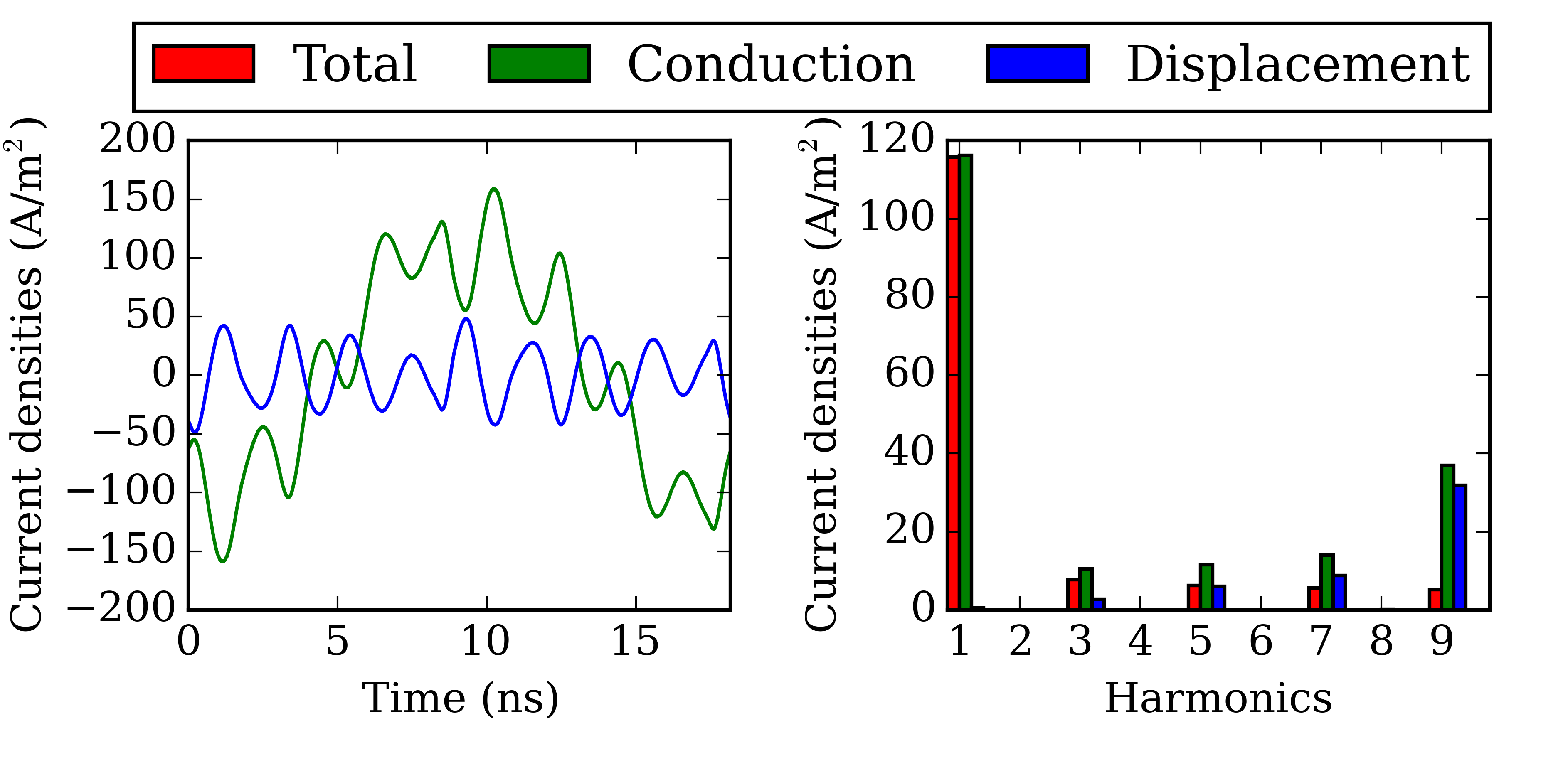}

		\caption{The conduction and displacement current density (left) for one RF-cycle and the corresponding Fourier spectra (right) in the center, at 7.5 mm ($\phi_0 = 300$ V).}
		\label{cur300}
	\end{center}
\end{figure}

Figure \ref{Curr} shows spatio-temporal plots of the conduction and displacement current density. The presence of energetic beam electrons causes a local enhancement of the conduction current density at distinct times within the RF period [Fig. \ref{Curr} (a)]. In order to ensure current continuity ($\nabla \cdot \vec{j}_{\rm tot} = 0$) the plasma must react to this local perturbation instantaneously. Consequently, a displacement current is generated, where the beam propagates [Fig. \ref{Curr} (b)]. This local displacement current is 180$^\circ$ out of phase with respect to the conduction current and compensates the perturbation caused by the presence of the beam electrons locally [Fig. \ref{Curr2}]. This can be understood as a local plasma parallel resonance (PPR) \cite{Annaratone,Ku,Ku2}. Physically, this displacement current is generated by the beam electrons themselves, since they move away from the expanding sheath edge. They leave behind a positive space charge, which the bulk electrons cannot instantaneously compensate. This space charge in turn causes an electric field, that first increases and then decreases, when the space charge is compensated by the cold bulk electrons. In this way current continuity is ensured in the presence of ballistic beam electrons based on this kinetic mechanism. In the voltage driven case, the displacement current compensates the conduction current only partially at higher odd harmonics of the driving frequency [Fig. \ref{Curr2}]. Thus, certain harmonics in the total current, and consequently, the PSR, are self-excited. (Even harmonics are not excited due to the symmetry of the discharge.)\\
Increasing the driving voltage amplitude ($\phi_0 = 300$ V) leads to a higher central plasma density and electron plasma frequency ($\omega_{\rm pe} \approx 9 \omega_{\rm RF}$ in the bulk). The cold bulk electrons can respond faster to the perturbation caused by the energetic beam electrons [Fig. \ref{ele300}]. They move back towards the expanding sheath edge on shorter timescales ($\tau \approx 2$ ns) and the above process can be repeated more often during one phase of sheath expansion. Consequently, the number of electron beams generated during one phase of sheath expansion increases [Fig. \ref{ele300} (a)] at higher driving voltage amplitudes. In the latter case, the kinetic interplay between low and high energetic electrons leads to the excitation of higher harmonics (9th harmonic) in the total current [Fig. \ref{VI_300V}] as well as in the current densities in the center of the discharge, which can be seen from the waveform and spectra given in Fig. \ref{cur300}.\\  
In summary, the mechanisms discussed here correspond to the first kinetic interpretation of resonance phenomena in CCRF plasmas. It is demonstrated that these effects cannot be understood in detail neither based on fluid models nor on global equivalent circuit models, since (i) local phenomena play a crucial role and (ii) a kinetic analysis is required to describe two groups of electrons propagating in opposite directions simultaneously. In principle, these kinetic mechanisms are present in any CCRF discharge (asymmetric or symmetric) and they are key processes that drive the electron heating dynamics. However, their experimental identification, particularly in the given symmetric case, is challenging due to the high temporal resolution required (on a sub-nanosecond timescale). 

Funding: This work has been supported by the German Research Foundation (DFG) within the frame of the Collaborative Research Centre TRR 87 and by the Hungarian Scientific Research Fund through the grant OTKA K-105476, NN-103150 and the Janos Bolyai Research Scholarship.

\newpage
%\section*{References}

\end{document}